\documentclass[prd,twocolumn,
superscriptaddress,altaffilletter,lengthcheck,tightenlines]{revtex4}

\usepackage{revsymb}
\usepackage{graphicx}

\newcommand{\be}{\begin{equation}}
\newcommand{\ee}{\end{equation}}
\newcommand{\ben}{\begin{eqnarray}}
\newcommand{\een}{\end{eqnarray}}
\newcommand{\n}{\label}
\newcommand{\no}{\noindent}
\newcommand{\ke}{$k$-essence }
\newcommand{\kf}{$k$-field }
\newcommand{\la}{\lambda}
\newcommand{\ga}{\gamma}

\begin{document}

\title{Interacting fluids generating identical, dual and phantom cosmologies}

\author{L.P. Chimento}
\email{chimento@df.uba.ar}
\affiliation{Departamento de F\'\i sica, Facultad
de Ciencias Exactas y Naturales, Universidad de Buenos Aires,
Ciudad Universitaria, Pabell\'on I, 1428 Buenos Aires,
Argentina,}

\bibliographystyle{plain}

%\maketitle

\begin{abstract}

We find the group of symmetry transformations generated by interacting fluids
in spatially flat Friedmann-Robertson-Walker (FRW) spacetime which links
cosmologies with the same scale factor {\it (identity)} or with scale factors
$a$ and $a^{-1}$ {\it (duality)}. There exists a duality between contracting
and superaccelerated expanding scenarios associated with {\it (phantom)}
cosmologies. We investigate the action of this symmetry group on
self-interacting minimally(conformally) coupled quintessence and $k$-essence
cosmologies.

\end{abstract}


\maketitle

%\vspace*{1cm}

\date{today}

%========================================================================
\section{Introduction}
%========================================================================

Symmetry transformations preserving the form of the spatially flat FRW
equations introduce an alternative concept of equivalence between different
physical problems \cite{si}. This means that a set of cosmological models are
equivalent when their dynamical equations are form invariant under the action
of that group. It suggests that any of this equivalent cosmologies can be used
to describe the present accelerated expansion of the universe. Hence, it turns
out to be interesting to investigate the consequences of this group.

Due to the additivity of the stress-energy tensor a cosmological model with
one fluid in flat FRW spacetime can be seen as a model of two interacting
fluid components. For instance a scalar field minimally coupled to gravity has
been described as a stiff fluid interacting with vacuum energy \cite{si}. The
stress-tensor of the tachyon field could be considered as the sum of two
components, one behaving like a pressureless dust and the other having a
negative pressure \cite{padma}. In this letter we investigate and extend these
results splitting the source in two components in a manner compatible with two
discrete symmetries of the Einstein equations in flat FRW spacetimes. These
symmetries are the structural invariance of the scale factor as a function of
the cosmological time, i.e., $\bar a(t)=a(t)$
\cite{ph},\cite{cataldo},\cite{2p},\cite{sen} and the duality between
expanding and contracting backgrounds $\bar a =1/a$
\cite{ph},\cite{2p},\cite{dabro1},\cite{lidsey}. In addition, the dual
transformation mapping contracting into superaccelerated expanding backgrounds
provides the link between a standard and a phantom cosmology
\cite{2p},\cite{dabro2}. A phantom source with sufficiently negative pressure
violates the weak energy condition but could describe adequately current
observations \cite{obs}. Other characteristics of phantom cosmologies have been investigated in \cite{todos}.

In section II we develop the interacting framework and illustrate it with
simple examples, such as, self-interacting minimally(conformally) coupled
quintessence field $\phi$($\psi$) and $k$-essence field $\varphi$. In section
III we present a linear transformations which preserve the form of the
dynamical equations and apply them to those scalar fields. In section IV the
conclusion are stated.

%%%%%%%%%%%%%%%%%%%%%%%%%%%%%%%%%%
\section{Interacting framework}
%%%%%%%%%%%%%%%%%%%%%%%%%%%%%%%%%%

The Einstein equations in the flat FRW spacetime with scale
factor $a$ and a perfect fluid read
\ben
\label{00}
3H^{2}=\rho,\\
\n{co}
\dot\rho+3H(\rho+p)=0,
\een

\no where $\rho$ is the energy density, $p$ the pressure and $H=\dot a/a$.
There are two independent Einstein equations for the three quantities $a,p,$
and $\rho $. Usually, the system of equations
(\ref{00}-\ref{co}) is closed with an equation of state $p=p(\rho)$.

We assume that $T_{ik}$ splits into two perfect fluid parts, $T_{ik} =
T_{ik}^1+ T_{ik}^2$, with $T_{ik}^{1,2} = (\rho_{1,2}+p_{1,2})
u_{i}u_{k}+p_{1,2}g_{ik}$, where $\rho_{1,2}$ and $p_{1,2}$ are the energy
density and the equilibrium pressure of fluids $1$ and $2$ respectively,
$u^{i}$ is the four-velocity. Therefore, Eqs. (\ref{00}-\ref{co}) become
\ben
\label{00i}
3H^{2}=\rho_1+\rho_2,\\
\n{coi}
\dot\rho_1+\dot\rho_2+3H(\rho_1+\rho_2+p_1+p_2)=0.
\een

\no The Eq. (\ref{coi}) shows the interaction between the fluid components
allowing the mutual exchange of energy and momentum. Consequently, there will
be no local energy-momentum conservation for the fluids separately. To
preserve the degree of freedom of the original system of equations
(\ref{00}-\ref{co}), we introduce an equation of state for each fluid
component $p_{1,2}=(\gamma_{1,2}-1)\rho_{1,2}$, where $\gamma_{1,2}$ are the
barotropic indexes of fluids $1$ and $2$ respectively. As the energy-momentum
tensor of the system as a whole is conserved, we assume an effective perfect
fluid description with equation of state $p=(\gamma-1)\rho$, where
$\gamma=(\gamma_1\rho_1+\gamma_2\rho_2)/\rho$ is the effective barotropic
index. For this effective perfect fluid the dynamical equations are identical
to (\ref{00}-\ref{co}).

The scalar field $\phi$ with energy density and pressure
\be
\n{rq}
\rho_\phi=\frac{1}{2}\dot\phi^2+V(\phi),
\quad p_\phi=\frac{1}{2}\dot\phi^2-V(\phi),
\ee

\no can be represented in terms of two interacting fluids namely
$\rho_1=\dot\phi^2/2$ and $\rho_2=V(\phi)$, with equations of state
$p_1=\rho_1$ and $p_2=-\rho_2$, meaning that $\gamma_1=2$
{\it (stiff matter)}
and $\gamma_2=0$ {\it (vacuum energy)}. Due to the interactions between the
two fluid components the energy-momentum tensor conservation of
the system as a whole is equivalent to the Klein-Gordon equation
\ben
\label{kgq}
\ddot{\phi}+3H\dot{\phi}+\frac{dV}{d\phi}=0,
\een

\no while $\gamma=2\rho_1/\rho_\phi$ is the effective barotropic index.

The conformal scalar field $\psi$ driven by the potential
\be
\n{fi4}
{\cal V}(\psi) = \lambda\psi^{4}+ {\cal V}_{0}, \qquad {\cal V}_{0}>0,
\ee

\no  is an interesting model, because this
potential has received much attention in the literature in connection
with the early inflationary epoch \cite{one}. This simplified model leads
to a final accelerated expansion phase retaining the essentials of
minimally coupled approaches. The energy density and the pressure of the
conformal scalar field
\ben
\n{rc}
\rho_\psi=\frac{1}{2}(\dot\psi+H\psi)^2+\lambda\psi^4+{\cal V}_0,\\
\n{pc}
p_\psi=\frac{1}{6}(\dot\psi+H\psi)^2+\frac{\lambda}{3}\psi^4-{\cal V}_0,
\een

\no can be represented as the sum of two interacting fluids namely
$\rho_1=(\dot\psi+H\psi)^2/2+\lambda\psi^4$ and $\rho_2={\cal V}_0$,
with equations of state $p_1=\rho_1/3$ and $p_2=-\rho_2$ representing
{\it (radiation)} $\gamma_1=4/3$ and {\it (vacuum energy)} $\ga_2=0$,
while $\ga=4\rho_1/3\rho_\psi$ is the effective barotropic index.
The energy conservation of the effective fluid is given by the
Klein-Gordon equation
\be
\n{kgc}
\ddot\psi+3H\dot\psi+\frac{1}{6}R\psi+4\lambda\psi^3=0,
\ee

\no whose first integral is
\be
\n{pi}
\frac{1}{2}(\dot\psi+H\psi)^2+\lambda\psi^4=\frac{b}{a^4},
\ee

\no where $b$ is an integration constant. After combining Eqs. (\ref{00}),
(\ref{rc}), (\ref{pi}) and integrating, we get the scale factors
\ben
\n{s4}
a_{c}^{\pm}=\left[\pm\sqrt{\frac{b}{{\cal V}_0}}
\sinh{\sqrt{\frac{4{\cal V}_0}{3}}\,t}\right]^{1/2}
\quad b>0\\
\n{s4'}
a_{c}=\left[\sqrt{\frac{-b}{{\cal V}_0}}
\cosh{\sqrt{\frac{4{\cal V}_0}{3}}\,t}\right]^{1/2}
\quad b<0 .
\een

\no Defining the conformal time $\eta=\int dt/a$ and the new field $\chi=\psi
a$, the Eq. (\ref{kgc}) becomes $\chi''+4\lambda\chi^3=0$, where $'\equiv
d/d\eta$. Its general solution can be expressed in terms of the Jacobi
functions. For $\lambda>0$, the qualitative aspects of $\psi$ are obtained
from the first integral (\ref{pi}), which reads $\chi'^2/2+\la\chi^4=b$.
Hence, $\chi$ oscillates between $-b^{1/4}\le\chi\le b^{1/4}$ and
$\psi=\chi/a$ becomes a decreasing oscillating function with a vanishing final
limit at the minimum of the potential. On the other hand, the bouncing
solution (\ref{s4'}) avoids the initial singularity.

For the \ke field $\varphi$ with Lagrangian ${\cal
L}=-U(\varphi)\,F(x)$, where $U(\varphi)$ is the potential, $F(x)$ is a
function of the kinetic term $x=g^{ik}\varphi_i\varphi_k$ and
$\varphi_i=\partial \varphi/\partial x^i$, we associate the
energy-momentum tensor of a perfect fluid. The energy density and
pressure of the $k$ field are
\be
\n{rk}
\rho_\varphi=U(\varphi)[F-2xF_x], \quad p_\varphi=-U(\varphi)F,
\ee

\no with $F_x=dF/dx$. They can be split as two interacting perfect
fluids such that $\rho_1=U(\varphi)F(x)$ and
$\rho_2=-2U(\varphi)xF_x$ with equations of state $p_1=-\rho_1$ and
$p_2=0$. They play the role of {\it (vacuum energy)} $\ga_1=0$ and {\it
(dust)} $\ga_2=1$. The energy-momentum tensor conservation
of the effective fluid is the \kf equation
\be
\n{kgk}
[F_x+2xF_{xx}]\ddot\varphi+3HF_x\dot\varphi+\frac{U'}{2U}[F-2xF_x]=0,
\ee

\no while $\gamma=\rho_2/\rho_\varphi$ is the effective barotropic index.

%%%%%%%%%%%%%%%%%%%%%%%%%%%%%%%%%%%%%%%%%%%%%
\section{Form invariance symmetry}
%%%%%%%%%%%%%%%%%%%%%%%%%%%%%%%%%%%%%%%%%%%%%

Here, we will find a symmetry transformation that preserves the
form of the system of equations (\ref{00i}) and (\ref{coi}).
To begin with, we observe that
the total energy density is form invariant under the linear
transformations
\begin{eqnarray}
%\begin{displaymath}
\left(\begin{array}{c}
\bar\rho_1\\
\bar\rho_2\\
\end{array}\right)
=
\left(\begin{array}{cc}
\alpha & 1-\beta\\
1-\alpha & \beta\\
\end{array}\right)
\left(\begin{array}{c}
\rho_1\\
\rho_2\\
\end{array}\right),
\n{tr}
%\end{displaymath}
\end{eqnarray}

\no that is, $\bar\rho=\bar\rho_1+\bar\rho_2=\rho_1+\rho_2=\rho$ for any
$\alpha$ and $\beta$. These form invariant transformations constitute a group
and they induce the transformations $\bar H=H$ or $\bar H=-H$ in Eq.
(\ref{00i}). The former leads to the {\it identity} $\bar a=a$ and the latter
to the {\it duality} $\bar a=1/a$. The duality between contracting and
superaccelerated expanding ( $H >0$ and $\dot H >0$ )
phases gives rise to a {\it phantom} transformation
\cite{2p}. Assuming equations of state for the interacting fluids,
$p_{1,2}=(\ga_{1,2}-1)\rho_{1,2}$ in the initial configuration and $\bar
p_{1,2}=(\bar\ga_{1,2}-1)\bar\rho_{1,2}$ in the final configuration, then the
conservation equation (\ref{coi}) remains form invariant when
\ben
\n{t+}
\alpha_{i}=\frac{\bar \ga_2-\ga_1}{\bar \ga_2-\bar \ga_1},
\quad \beta_{i}=\frac{\ga_2-\bar \ga_1}{\bar \ga_2-\bar \ga_1},
\quad \bar H=H,\\
\n{t-}
\alpha_{d}=\frac{\bar \ga_2+\ga_1}{\bar \ga_2-\bar \ga_1},
\quad \beta_{d}=-\frac{\ga_2+\bar \ga_1}{\bar \ga_2-\bar \ga_1},
\quad \bar H=-H.
\een

\no The duality $H\to -H$, acting on Eq. (\ref{co}) induces the
transformation $\rho+p\to -(\rho+p)$ and the associated matter violates
the weak energy condition. Such model could explain current observations and it is referred to as phantom cosmology \cite{phan}. When both $\rho$ and $p$ diverge the dual transformation trades a final big crunch by a final big rip. However, when $\rho$ is finite but $p$ diverges  \cite{Barrow}, the dual transformation interchanges this finite time singularity by other of the same kind. These sudden future singularities can occur even when the matter obeys $\rho>0$ and $\rho+3p>0$.

Below, we show the action of the form invariance symmetry group on the three
fields $\phi$, $\psi$ and $\varphi$. The associated effective perfect fluid
description will be used splitting each fluid as two interacting ones with
indexes $(\ga_1,\ga_2)$, i. e., $Q(\phi)\equiv (2,0)$, $C(\psi)\equiv (4/3,0)$
and $k(\varphi)\equiv (0,1)$. Also, we allow real or imaginary fields.

%%%%%%%%%%%%%%%%%%%%%%
%\subsection{$k\to Q$}
%%%%%%%%%%%%%%%%%%%%%%

{$k\to Q$}: Let us suppose that we are interested in seeking the set of
k-essence models having the same scale factor that the quintessence
ones. To this end we use Eqs. (\ref{rq}), (\ref{rk}), (\ref{tr}) and
(\ref{t+}). Then, we get
\ben
\n{t1}
\frac{1}{2}\dot\phi^2=\dot\varphi^2 UF',
\quad V=U\left[F-F'x\right]. \een

\no We are demanding that the last be an identity, i.e. it does not
impose any relation between the fields $\phi$, $\varphi$ and their
derivatives. Under this
requirement the second equation (\ref{t1}) translates into conditions
\be
\n{U=V}
V(\phi)=U(\varphi), \quad F-F'x=1.
\ee

\no The first equation indicates that both potentials are the same when
written as functions of cosmological time,
but different when written as functions of the individual fields.
Integrating the second Eq. (\ref{U=V}) we get $F=1+mx $
with $m$ an arbitrary integration constant. Inserting $F$ into
the first Eq. (\ref{t1}) we find the following relationship between the
fields $\phi$ and $\varphi$ \cite{q=k}
\begin{equation}
\phi=\sqrt{2m}\int\sqrt{U}d\varphi
\label{qk}.
\end{equation}

\no It shows the {\it identity} between quintessence and k-essence
models in flat FRW. For instance, the inverse square potential $U\propto
\varphi^{-2}$ correspond to an exponential potential $V\propto
e^{-\sqrt{2\phi}/\sqrt{m U_0}}$ \cite{ks} and
$U\propto\varphi^{2n}$ to $V\propto \phi^{2n/(n+1)}$.

For $k\to Q$ the dual transformation (\ref{tr}), (\ref{t-}) gives
\be
\n{t1-}
\frac{1}{2}\dot\phi^2=-\dot\varphi^2 UF', \quad V=U\left[F-3F'x\right].
\ee

\no Requiring that these equations be identities, we have
$V(\phi)=U(\varphi)$ and  $F=1-m x^{1/3}$.
Then, the first Eq. (\ref{t1-}) gives the transformation
rule for the kinetic terms
\ben
\n{tf-}
\dot\phi^2=\frac{2mU}{3}\dot\varphi^{2/3}.
\een

\no For constant potentials $V=U=V_0>0$, the first integrals of
Eqs. (\ref{kgq}) and (\ref{kgk}) can be written as
$\dot\phi^2=2\bar b/\bar a^6$ and $\dot\varphi^{2/3}=2ba^6$ where the
integration constants $\bar b$ and $b$ transform as
$\bar b=2mV_0b/3$ (see Eq. (\ref{tf-})). After solving the Eq. (\ref{00}),
we express the duality as
\be
\n{akq+}
a=\left[\pm\sqrt{\frac{2mb}{3}}\sinh{\sqrt{3V_0}t}\right]^{-1/3}
\to \bar a=\frac{1}{a},  \quad \bar b>0,
\ee
\be
\n{akq-}
a=\left[\sqrt{\frac{-2mb}{3}}\cosh{\sqrt{3V_0}t}\right]^{-1/3}
\to \bar a=\frac{1}{a},  \quad \bar b<0.
\ee

\no The phantom sector of the duality comprises the $(-)$ branch of Eq.
(\ref{akq+}) for $Q\to k$ and the branch (\ref{akq-}) for $k\to Q$. The
former represents a cosmology with a future big rip singularity at $t=0$
and a real(imaginary) $\varphi$ according to $m>0$($m<0$). The latter
describes a non singular cosmology with $\dot{\bar H}>0$ and an
imaginary $\phi$.

%%%%%%%%%%%%%%%%%%%%%%
%\subsection{$C\to Q$}
%%%%%%%%%%%%%%%%%%%%%%

{$C\to Q$}: We are going to find the relationships between
a conformal scalar field model driven by the
potential (\ref{fi4}) and a scalar
field one under an identical transformation. Using Eqs. (\ref{rq}),
(\ref{rc}-\ref{pc}), (\ref{tr}) and (\ref{t+}), we obtain
\ben
\n{t2}
\frac{1}{2}\dot\phi^2=\frac{1}{3}(\dot\psi+H\psi)^2+\frac{2}{3}
\lambda\psi^4,\\
\n{t2'}
V(\phi)=\frac{1}{6}(\dot\psi+H\psi)^2+\frac{1}{3}
\lambda\psi^4+{\cal V}_0,
\een

\no the potential $V(\phi)={\cal
V}_0+\dot{\phi}^2/4$
and the scale factors (\ref{s4}-\ref{s4'}). Besides, the potential
$V$ can be reconstructed as a function  of $\phi$ and
\be
\n{vqc}
\phi=\ln{\tanh{\sqrt{\frac{{\cal V}_0}{3}}\,t}},
\quad V={\cal V}_0\left[\cosh^2{\phi}-\frac{2}{3}\sinh^2{\phi}\right],
\ee
correspond to the solutions (\ref{s4}) and
\be
\n{vqc'}
\phi=-2i\arctan{e^{\sqrt{4{\cal V}_0/3}\,t}},
\quad V={\cal V}_0\left[\cos^2{i\phi}+\frac{2}{3}\sin^2{i\phi}\right],
\ee

\no to the solutions  (\ref{s4'}).

For $C\to Q$ the dual transformation (\ref{tr}), (\ref{t-}) gives
\ben
\n{t5}
\frac{1}{2}\dot\phi^2=-\frac{1}{3}(\dot\psi+H\psi)^2-\frac{2}{3}
\lambda\psi^4,\\
\n{t5'}
V(\phi)=\frac{5}{6}(\dot\psi+H\psi)^2+\frac{5}{3}
\lambda\psi^4+{\cal V}_0.
\een

\no These equations lead to the potential $V(\phi)={\cal
V}_0-5\dot{\phi}^2/4$, so comparing the first integral of the Eq. (\ref{kgq})
with Eqs. (\ref{pi}) and (\ref{t5}), we have $\dot{\phi}^2=2\bar b\bar a^4$
with $\bar b=-2b/3$.
After solving the Eq. (\ref{00}), we express the duality as
\be
\n{acq+}
a_c^{\pm}\to \bar a_c^{\pm}=\left(a_c^{\pm}\right)^{-1},  \quad \bar b<0,
\ee
\be
\n{acq-}
a_c\to \bar a_c=\left(a_c\right)^{-1},  \quad \bar b>0,
\ee

\no The phantom sector of the duality comprises the $(-)$ branch of Eq.
(\ref{acq+}) for $C\to Q$ and the branch (\ref{acq-}) for $Q\to C$. The former
represents a cosmology with a future big rip singularity at $t=0$, where
$\phi$ and $V$ are obtained from (\ref{vqc}-\ref{vqc'}) making the
substitution $\phi\to i\phi$. The latter describes a non singular cosmology
with $\dot H>0$.

%%%%%%%%%%%%%%%%%%%%%%%
%\subsection{$C\to k$}
%%%%%%%%%%%%%%%%%%%%%%%

{$C\to k$}: From Eqs. (\ref{rc}-\ref{pc}), (\ref{rk}), and
 (\ref{tr}-\ref{t+}), we have

\ben
\n{t3}
UF=-\frac{1}{6}(\dot\psi+H\psi)^2-\frac{1}{3}\lambda\psi^4+{\cal V}_0,\\
\n{t3'}
-2UxF'=\frac{2}{3}(\dot\psi+H\psi)^2+\frac{4}{3}\lambda\psi^4.
\een

\no These equations lead to $U={\cal V}_0$ and $F=1+m x^2$, which along with Eqs.
(\ref{00}) and (\ref{kgk}) give the general solution (\ref{s4}-\ref{s4'}). For
$m<0$, the singular solution (\ref{s4}) of this purely kinematics k-essence
model interpolates between radiation and de Sitter phases. For $m>0$ the non
singular solution (\ref{s4'}) bounces at $t=0$.

For $C\to k$ the dual transformations (\ref{tr}), (\ref{t-}) gives
\ben
\n{t3-}
UF=\frac{7}{6}(\dot\psi+H\psi)^2+\frac{7}{3}\lambda\psi^4+{\cal V}_0,\\
\n{t3'-}
-2UxF'=-\frac{2}{3}(\dot\psi+H\psi)^2-\frac{4}{3}\lambda\psi^4,
\een

\no with $U={\cal V}_0$ and $F=1+m x^{2/7}$. In this case
the solution of Eq. (\ref{00}) becomes Eqs. (\ref{s4}-\ref{s4'})
and the duality is expressed by Eqs.
(\ref{acq+}-\ref{acq-}).

%%%%%%%%%%%%%%%%%%%%%%%
%\subsection{$C\to C$}
%%%%%%%%%%%%%%%%%%%%%%%

$C\to C$: We complete this section investigating the duality between
conformally coupled quintessences. Duality among $Q\to Q$ or $k\to
k$ were studied in Refs. \cite{ph}-\cite{q=k}. From, Eqs. (\ref{rc}-\ref{pc}),
(\ref{tr}) and (\ref{t-}), we have
 \ben
\n{t4}
\frac{1}{2}(\dot{\bar\psi}+\bar H\bar\psi)^2+\bar\lambda\bar\psi^4
=-\frac{1}{2}(\dot\psi+H\psi)^2-\lambda\psi^4,\\
\n{t4'}
\bar{\cal V}_0 ={\cal V}_0+(\dot\psi+H\psi)^2+2\lambda\psi^4.
\een

\no Comparing Eqs. (\ref{pi}) and (\ref{t4}) we get $\bar b/\bar
a^4=-b/a^4=\bar b a^4$, so $\bar b=b=0$ and $\bar{\cal V}_0={\cal V}_0$
by Eq. (\ref{t4}).
The Eq. (\ref{00}) turns into $3H^2={\cal V}_0$ and $a=e^{H_0\,t}$,
with $H_0=\pm\sqrt{{\cal V}_0/3}$.
Integrating the Eq. (\ref{t4'}) we obtain the conformal fields
\be
\n{si}
\psi=\frac{H_0}{\sqrt{-2\la}[1-e^{H_0\,t}]},
\quad \bar\psi=-\frac{H_0}{\sqrt{-2\bar\la}[1-e^{-H_0\,t}]},
\ee

\no along with their transformation rule
\be
\n{tsi}
\bar\psi=\frac{\sqrt{-2\la}}{\sqrt{-2\bar\la}}
\left[\psi-\frac{H_0}{\sqrt{-2\la}}\right].
\ee

\no So, the duality induces a linear transformation group acting on
$\psi$ and there is no phantom sector because $\dot H=0$.

%\newpage

%=======================================================================
\section{Conclusions}
%=======================================================================

We have found a symmetry group generated by the additivity of the
stress-energy tensor and shown that the different forms of summing two
interacting fluid components induces two discrete symmetries in the Einstein
equations. They correspond to cosmologies with {\it identical} geometry or to
the {\it duality} between expanding and contracting backgrounds. The duality
between contracting and superaccelerated expanding scenarios connects standard
and {\it phantom} cosmologies.

The identical transformation relates two different cosmologies,
both with the same scale factor, so the choice of a particular
model to describe the evolution of the universe is not unique
because there are a variety of ways of combining two interacting
fluids with the same geometry. 
We highlight that in spatially flat FRW cosmologies, the dual transformation gives rise to linear relations between the components of the Einstein tensor, which read
$\bar G_0^0=G_0^0$, $\bar G_1^1=2G_0^0-G_1^1$ and so on for the remaining spatial components. Consequently, the conservation laws preserve their form i.e., $\bar\nabla_i\bar G_k^i=\nabla_i G_k^i=0$.
This analysis reveals that  no
matter how accurately  future experiments may come to determine
the cosmological observables there will always be a fundamental
uncertainty about which of the possible  models leading  to the
observed set of values has been chosen by Nature. This  uncertainty
could be avoided by considering other kind of frameworks as superstring field theory, M-theory or some  specific interaction between the parts in which the total fluid was divided. For instance, in Ref. \cite{id} it was shown that a particular interaction between both fluids reduces the linear transformations (\ref{tr}) to the identity restricting the form invariance symmetry.

We have shown that $k$-essence cosmologies generated by a linear kinetic
function and quintessence ones share the same scale factor.

The conformally coupled quintessence model we have studied is equivalent to a purely kinematics k-essence model generated by a quadratic kinetic term. It describes a
universe that begins to evolve as if it were {\it radiation} dominated
at early times and ends in a constant {\it vacuum energy} dominated phase.

For $Q\to k$, $C\to Q$ and $C\to k$, we have found a phantom symmetry between
contracting universes ending in a big crunch and expanding universes ending in
a big rip. In these examples, the transformation rule for the scalar fields we
have obtained is different than Wick rotation. In the case $C\to C$, the dual
transformation is so restrictive allowing only the duality between expanding
and contracting de Sitter geometries and inducing a linear relation among the
conformal scalar fields.

Finally, the linear transformations we have found can be
extended to the case of fluids with non constant barotropic indexes and we think
this is an interesting subject to be investigated in the future.

%=======================================================================
\section*{Acknowledgments}
%=======================================================================

This work was partially supported by the University of Buenos Aires and
Consejo Nacional de Investigaciones Cient\'\i tifas y T\'ecnicas under
Projects X224 and 02205.

%%%%%%%%%%%%%%%%%%%%%%%%%%%

\end{document}